\input harvmac

\vskip 1cm

 \Title{ \vbox{\baselineskip12pt\hbox{}}}
 {\vbox{
\centerline{ Correlation functions for chiral primaries  }
\centerline{ in D=6 Supergravity on $AdS_3 \times S^3$   }  }}

\centerline{$\quad$ {  Mihail Mihailescu }}
\smallskip
\centerline{{\sl  }}
\centerline{{\sl Brown  University}}
\centerline{{\sl Providence, RI 02912 }}
\centerline{{\tt mm@het.brown.edu}}
 \vskip .3in 
 
Six dimensional supergravities on $ADS_3 \times S^3$ present
interest due to the role they play in the $AdS/CFT$ correspondence.
The correspondence in this case states the equivalence between supergravity
on the given background and a still unknown conformal field theory. The conformal
field theory in question is expected to appear by deforming of the free conformal 
field theory on $S^N(T^4)$ in a way which preserves the superconformal symmetry. 
The purpose of this paper is to compute the first nontrivial corrections to the 
equations of motion for the chiral primary fields coming from supergravity. Using 
the methods already developed which involve nontrivial redefinitions of fields, 
we compute three-point correlation functions for scalar chiral primaries and 
notice similarities between their expressions and those obtained in the orbifold 
conformal field theory.

 
\Date{} 

\lref\zuchini{F.Bastianelli, R.Zucchini,{\it Three Point Functions of 
    Chiral Primary Operators in d=3, N=8 and d=6, N=(2,0) SCFT at large N,}
  hep-th/9907047}
\lref\florea{R.Corrado, B.Florea, R.McNees,{\it Correlation Functions of 
Operators and Wilson Surfaces in the d=6, (0,2) Theory in the large N limit,}
 Phys.Rev.D60(1999):085011, hep-th/9902153. }
\lref\roma{L.J.Romans, {\it Selfduality for interacting fields: 
    Covariant field equations for six-dimensional chiral supergravities,} 
   Nucl.Phys.B276(1986):71.}
\lref\sezgin{S.Deger, A.Kaya, E.Sezgin, P.Sundell, {\it Spectrum of D=6, N=4b
Supergravity on $AdS_3 \times S^3$,} Nucl.Phys.B536(1998): 110-140, 
    hep-th/9804166.}
\lref\leeseiberg{S.Lee, S.Minwalla, M.Rangamani and N.Seiberg,
{ \it Three point functions of Chiral operators in $D=4$, 
   $N=4$ SYM at large $N$,} Adv.Theor.Math.Phys.2(1998): 697-718,  
   hep-th/9806074.}
\lref\freedmanmathur{D.Freedman, S.Mathur, A.Mathusis and L.Rastelli,
   {\it Correlation functions in the \quad CFT(d)/AdS(d+1) correspondence,} 
  Nucl.Phys.B546(1999): 96-118, hep-th/9804058. }
\lref\malda{ J. Maldacena, { \it  The large N limit of superconformal 
                             field theories and supergravity,} 
                      Adv.Theor.Math.Phys.2(1998): 231-252, 
                       hepth/9711200.}
\lref\malstrom{ J. Maldacena, A. Strominger,{ \it AdS3 Black Holes and 
a Stringy Exclusion Principle,}  JHEP 9812 (1998) 005, hep-th/9804085.}
\lref\strovafa{ A.Strominger, C.Vafa, {\it Microscopic Origin of the 
    Bekenstein-Hawking Entropy, }Phys.\quad Lett. B379(1996): 99-104, 
hep-th/9601029.}
\lref\deboer{J. de Boer,{ \it Large N Elliptic Genus 
and AdS/CFT Correspondence }, JHEP 9905 (1999) 017, hep-th/9812240.}
\lref\martinec{F. Larsen, E. Martinec, 
{ \it U(1) Charges and Moduli in the D1-D5
         System, JHEP 9906 (1999) 019,}  hep-th/9905064.}
\lref\wittenseberg{N.Seiberg, E.Witten,{\it The $D1/D5$ System and 
  Singular CFT,} JHEP 9904(1999) 017, hep-th/9903224.}
\lref\jeram{A.Jevicki, S.Ramgoolam, {\it Non commutative geometry from the
    ADS/CFT correspondence,} JHEP 9904(1999):032, hep-th/9902059}
\lref\mijeram{A.Jevicki, M.Mihailescu, S.Ramgoolam,{\it Gravity from CFT on
 $S^N(X)$ CFT: Symmetries and Interactions,} hep-th/9907144.}
\lref\nastase{H. Nastase, D. Vaman, P. van Nieuwenhuizen, { \it Consistent nonlinear KK reduction of 11d supergravity 
     on $AdS_7\times S_4$ and self-duality in odd dimensions, } hep-th/9905075.} 
\lref\lee{S. Lee, {\it $AdS_5/CFT_4$ Four-point Functions of Chiral Primary Operators: Cubic Vertices, }hep-th/9907108}
\lref\aryt{G.Arutyunov, S.Frolov, { \it Some Cubic Couplings in Type IIB Supergravity on $AdS_5\times S^5$ and 
     Three-point Functions in $SYM_4$ at Large N,} hep-th/9907085}

\def\lpa{\nabla_{\nu} \nabla^{\nu}}
\def\lpaa{\nabla_{\rho} \nabla^{\rho}}
\def\lps{\nabla_{b} \nabla^{b}}
\def\lpss{\nabla_{c} \nabla^{c}}
\def\ld{\nabla_{\nu}}
\def\lu{\nabla^{\nu}}

\def\ur{U^{r~I_{1}}}
\def\urp{U^{r~I_{2}}}
\def\uo{U^{r}}
\def\ua{U^{5~I_{2}}}
\def\phr{\phi^{5r~I_{1}}}
\def\php{\phi^{5r~I_{2}}}
\def\ph{\phi^{5r}}
\def\na{N^{I_{2}}}
\def\sr{s^{r~I}}
\def\sp{s^{r~I_{1}}}
\def\sq{s^{r~I_{2}}}
\def\sa{\sigma^{I}}
\def\ra{\longrightarrow}
\def\ca{C(I,I_{1},I_{2})}
\newsec{Introduction}
The correspondence $\malda$ between CFT and compactified supergravity 
on AdS spaces is a topic under vigorous current investigation. The first 
check of the correspondence was done by comparison of the spectra of 
chiral primaries in both theories and was followed by comparison
of three point interactions. Such comparison involve computation of 
three point Greens functions in Yang-Mills theory (without interactions)
and then, the evaluation of the analogue three point vertices in 
compactified supergravity $\freedmanmathur$,$\leeseiberg$,$\lee$,$\aryt$. 
An agreement of the two (even for the range of couplings where it should not hold)
is due to some rather nontrivial nonrenormalization theorems. 
\par
One of the most interesting case of Maldacena's conjecture is the one
of IIB string theory compactified on a $K3$ or $T^4$ manifold $\malstrom$.
Here, a distribution of five branes and one-branes gives a black string
whose near horizon geometry is $AdS_3 \times S^3 \times M^4$. Further 
compactification on $S^1$ produces a five-dimensional black hole. Its 
entropy is nicely accounted $\strovafa$ by a conformal field theory
with target space $S^N(T^4)$ or $S^N(K3)$. This orbifold conformal 
field theory was recently studied in a number of works $\deboer$, 
$\martinec$, $\wittenseberg$. In $\jeram$, a suggestion that it originates 
in a non-commutative space-time was given, while the nonlinear 
aspects of this conformal field theory were discussed in $\mijeram$. 
In particular, a construction of chiral primary operators was explicitely 
given and their operator products evaluated. This gave specific answers 
for the three point correlation functions.
The orbifold framework allowed for  a discussion of the corresponding
spectrum generating algebra which seems to govern the dynamics of the theory.
It is clearly relevant to perform the analogous calculations at the
supergravity level. In this case, the relevant $6D$ supergravity involves 
nontrivial multiplets which have their origin in compactification
on $K3$ or $T^4$. Furthermore, the precise correspondence is actualy not
known, in particular it is likely that a deformed conformal field theory
might ultimately be relevant. Even though some of the deformations 
have been mentioned, explicite calculations were not done yet. 
\par
In this paper, detailed computations of cubic interactions among 
chiral primaries of $6D$ supergravity will be presented. The computation 
involves the standard elements of previous works, namely nonlinear field 
redefinitions, which then leads to three point functions of chiral 
operators. The results presented at the end of section 4 have some 
similarities in form with the previous examples done in the literature 
(which all have a universal form in dependence on the chiral quantum 
numbers $l$) $\florea$, $\zuchini$. In our case, we observe for one 
sequence of operators a deviation from the universal form
and a particular quadratic dependence on the quantum numbers $l$. It is a 
very interesting question to ask if a deformation of the free orbifold CFT
would give an identical result. 
\par
The paper is organized as follows: 
in section 2 we give following $\roma$, $\sezgin$ the basic notation 
and definition of $6D$  supergravity with the discussion of chiral 
primaries. In section 3 the action is expanded to cubic order and the 
nonlinear field redefinitions are determined. Section 4 discusses the 
corresponding correlation functions. Some comments on future work are 
made in the conclusions.

\newsec{Supergravity in $D=6$}
In this section, we review  following $\roma$, $\sezgin$ the equations of 
motion of $D=6$ supergravity of interest for us in determining the scalar 
chiral primaries. We are interested in the equation of motion for metric 
$g_{MN}$, 2-forms $B^{i}_{MN}, B^{r}_{MN}, i=1..5, r=1..n$ and scalars.
The scalars are packaged in a $SO(5,n)$ matrix $V_{I}^{~J}, I,J=1..5+n$ 
giving a parametrization of the coset ${SO(5,n) \over {SO(5)\times SO(n)}}$.
The integer $n$ depends on the type of supergravity we study in $6D$: for 
$\it{N}=4b$ supergravity, coming from a compactification of type $IIB$ 
supergravity on $K3$, $n=21$ and for $\it{N}=8$ supergravity, coming from 
compatification of the same $10D$ theory on $T^4$, $n=5$. As long as we 
work with the metric, the scalars and two-forms only, we can use $n$ 
unspecified. We will use the following notation for indices: 
$M,N=0..5$ are $D=6$ coordinates, $\mu, \nu=0..2$ are $AdS_{3}$ coordinates 
and $a,b=1..3$ are $S^{3}$ coordinates. We will also use $G^{I}=dB^{I}$.
\par
We construct the following quantities which will be used in writing the 
equations of motion in a compact form:
\eqn\edeff{\eqalign
{ & dV V^{-1}=\pmatrix{ & Q^{ij} & \sqrt{2} P^{is} \cr
                        & \sqrt{2} P^{jr} & Q^{rs} \cr
                       }, \cr
  & \cr					   
  & D_{M}P_{N}^{ir} = \nabla_{M} P_{N}^{ir}-Q_{M}^{ij} P_{N}^{jr}-
     Q_{M}^{rs} P_{N}^{is}, \cr
  & H^{i}=G^{I} V_{I}^{~i},~~~~ H^{r}=G^{I} V_{I}^{~r},  \cr     
}}
Then, the $D=6$ supergravity equations of motion are:
\eqn\edeqq{\eqalign
{ & R_{MN}=H^{i}_{MPQ}H^{i~~PQ}_{N}+H^{r}_{MPQ}H^{r~~PQ}_{N}+
          2P_{M}^{ir}P_{N}^{ir}, \cr
  & D^{M}P_{M}^{ir}-{\sqrt{2} \over 3}H^{i}_{MNP}H^{r~~MNP}=0, \cr
  & *H^{i}=H^{i},~~~~ *H^{r}=-H^{r}. \cr
}}
We consider a vacuum solution which is maximally supersymmetric in $D=6$ and
has the geometry of $AdS_{3}\times S^{3}$ with radius and cosmological
constant equal to one. It consists in the corresponding metric 
$\bar{g}_{MN}$ of the direct product of the two spaces and the following 
ansatz for the forms and scalars: $\bar{V}=I_{5+n}~$, 
 $\bar{G}^{5}_{\mu \nu \rho}=\epsilon_{\mu \nu \rho}$,  
$\bar{G}^{5}_{abc}=\epsilon_{abc}$ and all the other $\bar{G}$'s set to zero. 
In the previous equations $\epsilon_{\mu \nu \rho}$ and $\epsilon_{abc}$ are
the volume forms on  $AdS_{3}$ and $S^{3}$ respectively such that,
$\epsilon_{\mu \nu \rho a b c}=\epsilon_{\mu \nu \rho}~\epsilon_{abc}$ is the
volume form in six dimensions. We will recover later by scaling arguments 
the radius dependence.    

\newsec{Scalar Chiral primaries}
We parameterize the scalar manifold by introducing scalar fields 
$\phi^{ir}$ representing a $5 \times n$ matrix $\Phi$ which gives 
the fluctuations of $V$ away from identity. We write all the 
fields in second order in fluctuations:
\eqn\fluct{\eqalign
{ & g_{MN}=\bar{g}_{MN}+h_{MN}, \cr
  & G^{I}=\bar{G}^{I}+g^{I}, \cr
  & \cr
  & V=\pmatrix{ & I_{5}+{1 \over 2} \Phi \Phi^{T} & \Phi \cr
                & \Phi^{T} &   I_{n}+{1 \over 2} \Phi^{T} \Phi \cr}, \cr
}}
Then, we write  the quantities defined in $\edeff$ in this order and  
we obtain:
\eqn\fluctsec{\eqalign
{ & Q^{ij}={1 \over 2}(\phi^{ir} d\phi^{jr}-\phi^{jr} d\phi^{ir}),\cr
  & Q^{rs}={1 \over 2}(\phi^{ir} d\phi^{is}-\phi^{is} d\phi^{ir}),\cr
  & P^{ir}={1 \over \sqrt{2}}d\phi^{ir}, \cr
  & H^{i}=g^{i}+g^{r} \phi^{ir} + {1 \over 2}\bar{G}^{5}\phi^{5r}\phi^{ir}, 
   i \neq 5 , \cr
  & H^{5}=\bar{G}^{5}+g^{5}+g^{r} \phi^{5r} + {1 \over 2}\bar{G}^{5}\phi^{5r}\phi^{5r}, \cr
  & H^{r}=g^{r}+\bar{G}^{5}\phi^{5r}+g^{i}\phi^{ir}, \cr 
}}
We focus now, on scalar chiral primaries only. As it was already shown in
$\sezgin$,  they come from two sources: for $r=1..n$ from the equation 
of motion for the two-forms $B^{r}$ and scalars $\phi^{5r}$, and for the 
last one from the equation of graviton and the two-form $B^{5}$. We  
also use the parametrization of the two-forms given in \sezgin. The forms are 
decomposed in spherical harmonics of $S^{3}$ and we use the gauge 
invariance and the allowed diffeomorphism invariance to reduce the 
fluctuations to physical ones:
\eqn\parm{\eqalign
{ & h_{ab}=N \bar{g}_{ab},~~~~ h_{\mu a}=0, ~~~~
h_{\mu \nu}=M \bar{g}_{\mu \nu}+h'_{(\mu \nu)}, \cr
  & g^{I}_{MNP}=3\partial_{[M} b^{I}_{NP]}, \cr
  & b^{I}_{\mu \nu}=\epsilon_{\mu \nu \rho}X^{I~\rho},~~~~
 b^{I}_{\mu a}=0,~~~~ b^{I}_{abc}=\epsilon_{abc}\partial^{c}U^{I},\cr
}}   
where $h'_{(\mu \nu)}$ is the traceless part of the graviton fluctuation 
on $AdS_{3}$.  
\par
Let us start with the first category of chiral primaries and solve the 
linearized equations leading to those 
(see appendix):
\eqn\moteq{(\nabla^{\mu}\nabla_{\mu}+\nabla^{a}\nabla_{a})\phi^{5r}+
   4(\nabla^{\mu}\nabla_{\mu}-\nabla^{a}\nabla_{a})U^{r}=
   4\nabla^{\mu}(\nabla_{\mu}U^{r}-X^{r}_{\mu})+Q^{r}_{1},
}
from the equation of motion for scalars $\ph$,
\eqn\motqe{ \nabla_{a}X^{r}_{\mu}-\nabla_{a}\nabla^{\mu}U^{r}=Q^{r}_{2~a\mu},
}
and
\eqn\moeqt{(\nabla^{\mu}\nabla_{\mu}+\nabla^{a}\nabla_{a})U^{r}+2\phi^{5r}
     =\nabla^{\mu}(\nabla_{\mu}U^{r}-X^{r}_{\mu})+Q^{r}_{3}, 
}
from the $\mu \nu a$ and $abc$ component of the antiself-duality equation for 
the two-forms $B^{r}$. 
We expand the scalars $\ph$ and those used to parametrize the fluctuations of 
two-forms in the harmonic basis of $S^{3}$ and we get after solving the 
constraint $\motqe$, a system of equations , $\moteq$ and $\moeqt$,  
which has to 
be diagonalized. We arrive at the following definitions: for the chiral 
primaries $s^{r~I}={1 \over 4(l+1)}(\phi^{5r~I}+2(l+2)U^{r~I})$ and for the
other scalar $t^{r~I}={1 \over 4(l+1)}(\phi^{5r~I}-2lU^{r~I})$,
where $I$ refers from now on to the harmonic index on $S^{3}$ and $l$ refers 
to the integer entering the Casimir of that representation: $l(l+2)$. We also 
have, if we put to zero all the fields except the chiral primaries:
$\phi^{5r~I}=2ls^{r~I}, U^{r~I}=s^{r~I}$.

\par

For the second category, those coming from the graviton and one of the 
two-forms, we have the following equations at the linearized level:
\eqn\mete{-{1 \over 2}(\nabla^{\mu}\nabla_{\mu}+\nabla^{a}\nabla_{a})N-
           {1 \over 6}\nabla^{a}\nabla_{a}(3M+N)=-4N+
            4\nabla^{a}\nabla_{a}U^{5}+Q^{\sigma}_{1},}
and
\eqn\mtee{ (3M+N)_{;(ab)}=Q^{\sigma}_{2~(ab)},} 
coming from the $ab$ component of the Einstein equation; $\mete$ is the part 
proportional to $\bar{g}_{ab}$ and $\mtee$ is the symmetric traceless part 
$(ab)$;      
\eqn\metef{ \nabla_{a}X^{5}_{\mu}+\nabla_{a}\nabla^{\mu}U^{5}+
  Q^{\sigma}_{3~a\mu }=0,
}
and
\eqn\metfe{(\nabla^{\mu}\nabla_{\mu}+\nabla^{a}\nabla_{a})U^{5}-2N
     =\nabla^{\mu}(\nabla_{\mu}U^{5}+X^{5}_{\mu})+Q^{\sigma}_{4} 
}
from the $\mu \nu a$ and $abc$ component of the self-duality equation
for the two-form $B^{5}$. In this equations $Q^{r}_{1,2,3}, 
Q^{\sigma}_{1,2,3,4}$ represent the second order corrections to the 
equations and they are set to zero when we determine the linearized expressions 
for chiral primaries. We solve again the constraints $\mtee$ and $\metef$, and 
we obtain a system of equations $\mete$ and $\metfe$ which is diagonalized 
by the following expressions: the chiral primaries given by 
$\sigma^{I}={1 \over 4(l+1)}(-N^{I}+2(l+2)U^{5~I}),$ and another nonchiral 
primary scalar $\tau^{I}={1 \over 4(l+1)}(N^{I}+2lU^{5~I})$. If we set to 
zero all fields, but the chiral primaries we obtain the following expressions:
$N^{I}=-2l\sigma^{I}, U^{5~I}=\sigma^{I}$. We solve also the linearized
$\mu a$ part of the graviton equation and we obtain that 
$h'^{I}_{(\mu \nu)}=-{4 \over {l+1}}\nabla_{(\mu}\nabla_{\nu)}\sigma^{I}$.
We note that in the above equations, namely at the linearized level,
$h'^{I}_{(\mu \nu)}$ does not enter. It apears only in the second order
variation of equations of motion (in $Q$'s), for our purpouses being enough 
to compute only its first order expresssion in term of chiral primaries. 

\newsec{The action in $AdS_{3}$ for chiral primaries}
We compute in this section the second order corrections to the 
equations of motion for chiral primaries. We notice that some of the 
second order terms involve derivatives on $AdS_{3}$ of chiral primaries. 
Because a consistent field theory believed to exist for the chiral primaries
requires the absence of these terms, we will redefine the chiral primaries 
in such a way that the new terms cancel the problematic terms. These 
redefinitions of chiral fields are also consistent to the fact that we
named the chiral primaries after solving linearized equations only. It
will be intersting to test if using the full nonlinear expression for the 
supersymmetry algebra and the conditions for a field to be chiral we could 
confirm our redefinitions of fields.
\par
Let us write the expressions for the corrections using partial results
computed in appendix 1. We start with the corrections given in $Q^{r}$'s:
\eqn\corre{\eqalign
{& Q^{r}_{1}=\nabla_{\mu}(h^{\mu \nu}\nabla_{\nu} \phi^{5r})+
             \nabla_{a}(h^{ab} \nabla_{b} \phi^{5r})-
              {1 \over 2} \nabla^{P}(h_{M}^{~M}) \nabla_{P} \phi^{5r}+ \cr
    &~~~ +4(h_{\mu}^{~\mu} \lpa U^{r}-h_{a}^{~a} \lps U^{r}+
            2\phi^{5r}(\lpa U^{5}+\lps U^{5}-2N))+ \cr
    &~~~ +4(\lpa U^{5} \lpaa U^{r}+
            \lps U^{5} \lpss U^{r}+
            2 \nabla_{b} \nabla_{\nu} U^{5}\nabla^{b} \nabla^{\nu} U^{r}),\cr
 & Q^{r}_{2~ a \mu}=2 \nabla_{a} \nabla_{\mu} U^{5} \phi^{5r}
       +\nabla_{a} \nabla_{\mu} U^{r} (h_{\mu}^{~\mu}-{1 \over 2}h_{M}^{~M})
        +h_{a}^{~b} \nabla_{b} \nabla_{\mu} U^{r}-
          h_{\mu}^{~\nu} \nabla_{a} \nabla_{\nu}U^{r}, \cr
 &  Q^{r}_{3}=(\lpa U^{5} - \lps U^{5}) \phi^{5r}-
          ({1 \over 2} h_{M}^{~M} - h_{a}^{~a}) (\lps U^{r} + \phi^{5r}).\cr} 
}
\par
Next, we use the expansion of fields in sperical harmonics and compute the
correponding corrections for each $I$. For definitions and derivations for 
spherical harmonics see appendix 2. Here, we only repeat a few definitions:
$\Delta=l(l+2)$ are up to a minus sign the eigenvalues of the sphere laplacian,
$\delta=l(l-2)$ are related to the masses of the chiral primaries, $C(I)$ is  
the normalization constant of the $I$-th  spherical harmonic and is obtained
from the integral over  $S^{3}$ of the square of it, and $C(I,I_{1},I_{2})$
is the integral of the corresponding product of three spherical harmonics.
\par
We solve first the constraint $\motqe$  and  from it we obtain: 
\eqn\cerr{\eqalign
{& \nabla_{\mu} U^{r~I}-X_{\mu}^{r~I}=
            {C(I,I_{1},I_{2}) \over {\Delta C(I)}}(-2l_{1}
      (\Delta+\Delta_{2}-\Delta_{1})s^{r~I_{1}} \nabla_{\mu}\sigma^{I_{2}}+\cr
&~~~  +(\Delta+\Delta_{1}-\Delta_{2})(-l_{2} \nabla_{\mu}s^{r~I_{1}}
     \sigma^{I_{2}}+{1 \over 2} \nabla_{\nu}s^{r~I_{1}} h_{\mu}^{I_{2}~\nu})).\cr
}}  

We expand the equations $\moeqt$ and $\moteq$ in spherical harmonics and  for 
each $I$ we sum them with the followig weights: $\moeqt$ with 
${(l+2) \over 2(l+1)}$
and $\moteq$ with ${1 \over 4(l+1)}$. In the end, we obtain the following 
equation for $\sr$ chiral primaries including second order corrections:
\eqn\seqr{\eqalign
{ & (\lpa -l(l-2))s^{r~I}={1 \over 4(l+1)}(2(l+2)\nabla^{\mu}
   (\nabla_{\mu} U^{r~I}-X_{\mu}^{r~I})+\cr
 &~~~+{C(I,I_{1},I_{2}) \over C(I)}(\nabla_{\mu}(\nabla_{\nu}\phi^{5r~I_{1}} 
     h^{I_{2}~\mu \nu})-\Delta_{1} \phi^{5r~I_{1}}N^{I_{2}}-
   \nabla_{\mu}\phi^{5r~I_{1}}\nabla^{\mu}N^{I_{2}}+ \cr 
 &~~~+4((-\lpa + 3\Delta_{1})U^{r~I_{1}} N^{I_{2}}-
   4\phi^{5r~I_{1}}N^{I_{2}}+(-2\phr+
   \Delta_{1}\ur)\Delta_{2} \ua+\cr
 &~~~+(\lpa \ur+2 \phr)\lpa \ua+(\Delta_{1}+\Delta_{2}-\Delta)\ld \ur \lu \ua)+\cr
 &~~~  +2l(\phr(\lpa+\Delta_{2})\ua+2(-\Delta_{1}\ur+\phr)\na))).\cr
   }}
We use now the expression for the fields 
 $\phi^{5r}, U^{r},U^{5},N, h_{\mu\nu}..$ in term
of chiral primaries $\sr$ and as it is expected, we obtain equations which have 
in the right hand side terms involving derivatives $\ld \sr$. These terms 
suggest that the definition of chiral primary fields is changed at the 
quadratic level. We remove the unwanted terms by redefinitions and we 
end up with the following rules which are applied in the right hand side 
of $\seqr$.  We consider $s,\sigma$ as generic fields which 
at the linearized level solve $(\lpa-l_1(l_1-2))s=0$ and 
$(\lpa-l_2(l_2-2))\sigma=0$ respectively, and the rules are valid if
the corrected field solves at linearized level $(\lpa-l(l-2)l)...=0$.
\eqn\rules{\eqalign
{ & \ld (\lu s \sigma) \ra {1 \over 2} (\delta+\delta_{1}-\delta_{2})
         s \sigma, \cr
  & \lu s \ld \sigma \ra {1 \over 2} (\delta-\delta_{1}-\delta_{2})
         s \sigma, \cr	
  &	\nabla_{\mu} (\ld s \nabla^{\mu} \lu \sigma) \ra {1 \over 4} 
  (\delta+\delta_{2}-\delta_{1})(\delta-\delta_{1}-\delta_{2})s \sigma. \cr
  }}
\par
We mention here, that the above rules are valid only for second order 
corrections and they have to be changed or replaced for the higher orders. 
Plugging in $\seqr$ the expression for $X_{\mu}^{r~I}$ obtained from equation 
$\cerr$ and using then $\rules$ to remove the derivatives terms we will obtain 
the following equations :
\eqn\sfin{
(\lpa -l(l-2))s^{r~I}={\ca \over C(I)}{{\alpha \alpha_{1} \alpha_{2}
 (\Sigma-2) \Sigma (\Sigma+2)} \over {4 l (l+1) (l_{2}+1)}}
 s^{r~I_{1}}\sigma^{I_{2}},
 }
where  $\alpha=l_{1}+l_{2}-l$, $\alpha_{1}=l+l_{2}-l_{1}$, 
 $\alpha_{2}=l+l_{1}-l_{2}$ and $\Sigma=l_{1}+l_{2}+l$.
\par
We will work now, in a similar fashion the quadratic corrections to
the equations of motion for the chiral primary  coming from the metric and one
of the self-dual forms, $\sigma$. There are two kinds of quadratic corrections:
those which are quadratic in $\sr$ and those which are quadratic in $\sa$. 
We work each correction separately. We start with those coming from the first
category (we will denote both in the same way, $Q$'s):
\eqn\scorr{\eqalign
{& Q^{\sigma}_{1}=4 \lps \uo \ph + 2 \ph \ph + 
            2 ( \lps \uo +\ph)^{2} + {2 \over 3}(\nabla_{b}\ld \uo)^{2}+
			{1 \over 3}(\nabla_{b}\ph)^{2}, \cr
 & Q^{\sigma}_{2~(ab)}=8\nabla_{(a}\lu \uo \nabla_{b)}\ld \uo-
             2\nabla_{(a}\ph \nabla_{b)}\ph, \cr
 & Q^{\sigma}_{3~a\mu}=2\nabla_{a}\nabla_{\mu} \uo \ph, \cr
 & Q^{\sigma}_{4}=(\lpa-\lps)\uo \ph-{1 \over 2}(3M+N). \cr
 }}
We compute from $\mtee$ and $\metef$ the quadratic corrections to
$M^{I}$ and $X^{I~5}_{\mu}$ and plug them back in the system of equations 
formed from $\mete$ and $\metfe$. We sum then the two equations with the 
appropriate factors: $-{1 \over 4(l+1)}$ for $\mete$ and  
${(l+2) \over 2(l+1)}$ for $\metef$, and the result is the 
linearized equation for $\sa$ and its quadratic corrections. 
Expressing all the fields in term of $\sr$ and $\sa$ we obtain for the
constraints:
\eqn\sscorr{\eqalign
{ & 3M^{I}+N^{I}={{3 \ca} \over {2\Delta(\Delta-3)C(I)}}
({1\over 4}(\Delta_{1}+\Delta-\Delta_{2}) (\Delta+\Delta_{2}-\Delta_{1})+
{\Delta \over 6}(\Delta_{1}+\Delta_{2}-\Delta))\times \cr
  &~~~\times 8 (\lu \sp \ld \sq-l_{1}l_{2}	\sp \sq)+.., \cr
  & \lu (X_{\nu}^{I~5}+\ld U^{I~5})={{-2l_{2} \ca} \over {\Delta C(I)}} 
                 (\Delta_{1}+\Delta-\Delta_{2})\lu(\ld \sp \sq)+..,\cr
}}				 
where the dots represent the other kind of corrections, due to $\sa$. 
The final equation is:
\eqn\sccor{\eqalign
{& (\lpa -l(l-2))\sa={1 \over 4(l+1)}((-{\Delta \over 3}-l-2)
    (3M^{I}+N^{I})+2(l+2) \times  \cr
 &~~~ \times \lu (X_{\nu}^{I~5}+\ld U^{I~5})+
   {2 \ca \over C(I)}(-4\Delta_{1}\ur \php+2 \phr \php + \cr
 &~~~+2(-\Delta_{1}\ur+\phr)(-\Delta_{2}\urp+\php)+
(\Delta_{1}+\Delta_{2}-\Delta) \times \cr
 &~~~ \times ({1 \over 3}\lu \ur \ld \urp +
 {1 \over 6}\phr \php)+(l+2)(\lpa+\Delta_{1})\ur \php))+..\cr
}}				  
Rewriting all the fields in term of chiral primaries and using the given
rules $\rules$ to remove the quadratic corrections with derivatives we
obtain after symetrization in $I_{1,2}$ the following result:
\eqn\sfinr{
(\lpa -l(l-2))\sa={{\alpha \alpha_{1}\alpha_{2} (\Sigma-2)\Sigma(\Sigma+2)} \over
            {8(l-1)l(l+1)}}{\ca \over C(I)}\sp \sq+...
			}

We turn now to the quadratic corrections due only to $\sa$:
\eqn\scorr{\eqalign
{& Q^{\sigma}_{1}=-{1 \over 3}\lps (N^{2})-{1 \over 2}(N_{\mu}h^{\mu \nu})_{;\nu}
      -{1 \over 12} \lps (h^{\mu \nu}h_{\mu \nu})+
	  {1 \over 12}\nabla^{b}h^{\mu \nu} \nabla_{b}h_{\mu \nu}+\cr
 &~~~+{1 \over 12}\nabla^{b}N \nabla_{b}N +6N^{2}-8N \lps U^{5}+
	  +2(\lps U^{5})^{2}+{2 \over 3} (\nabla_{b}\ld U^{5})^{2}, \cr
 & Q^{\sigma}_{2~(ab)}={1 \over 2}((N^{2})_{;(ab)}+(h^{\mu \nu}h_{\mu \nu})_{;(ab)}-
              \nabla_{(a}h^{\mu \nu} \nabla_{b)}h_{\mu \nu}-\nabla_{(a}N \nabla_{b)}N)+\cr
 &~~~+8\nabla_{(a}\ld U^{5}\nabla_{b)}\lu U^{5},\cr
 & Q^{\sigma}_{3~a\mu}=-N \nabla_{a}\nabla_{\mu}U^{5}-\nabla_{a}\ld U^{5} h^{\nu}_{\mu}, \cr
 & Q^{\sigma}_{4}=2N\lps U{5}-{11 \over 4}N^{2}+{1 \over 4}h^{\mu \nu}h_{\mu \nu}
          -{1 \over 2}(3M+N). \cr
 }}
In the same way as before, we obtain the following corrected equations 
of motion after using the expressions of the $D=6$ fields in terms of the
chiral primaries, of the rules $\rules$ and of symmetrization in the 
spherical indices $I_{1}, I_{2}$:
\eqn\sfins{
(\lpa -l(l-2))\sa=...+{{\alpha \alpha_{1}\alpha_{2} (\Sigma-2)\Sigma(\Sigma+2)} \over
            {16(l-1)l(l+1)(l_{1}+1)(l_{2}+1)}}(l^{2}+l_{1}^{2}+l_{2}^{2}-2)
			{\ca \over C(I)}\sigma^{I_{1}} \sigma^{I_{2}},
			}
where by $...$ we represent the already computed part, namely, the RHS of $\sfinr$.
\par 
It is now possible to write the action compatible with the determined equations $\sfin$, $\sfinr$ and $\sfins$ whithout being necessary to compute from somewhere else the quadratic terms. The $D=3$ lagrangian, 
up to normalization constants, is:
\eqn\lagr{\eqalign
{& L_{chiral}=4l(l+1)C(I)((\partial \sr)^{2}+l(l-2)(\sr)^{2})
           +4l(l-1)C(I) \times \cr
 &~~~ \times ((\partial \sa)^{2}+l(l-2)(\sa)^{2})+
{{\alpha \alpha_{1} \alpha_{2}
 (\Sigma-2) \Sigma (\Sigma+2)} \over (l+1)} \ca \sa \sp \sq+ \cr
&~~~+{{\alpha \alpha_{1} \alpha_{2}
 (\Sigma-2) \Sigma (\Sigma+2)} \over {6(l+1)(l_{2}+1) (l_{2}+1)}}
 (l^{2}+l_{1}^{2}+l_{2}^{2}-2) \ca \sigma^{I}\sigma^{I_{1}}\sigma^{I_{2}}. \cr
 }}
\newsec{Correlation functions}
In this section, we derive the needed constant required in equation $\lagr$ 
from the higher dimensional theory and compute the CFT correlation functions. 
We start with a system of $Q_{1}$ $D1$-branes 
and $Q_{5}$ $D5$-branes in type IIB string theory. We  review the 
decoupling limit following $\malda$, $\malstrom$ and deduce from it the 
constant required in the equations of motion of $D=6$ supergravity. The 
following constants have the known definitions: $g$ is the string coupling,
$\alpha'$ is the inverse of the string tension, $v$ is the volume of 
$T^{4}$ and $g_{6}=g / {\it v}^{1 \over 2} $ is the $D=6$ coupling constant.
The $D=10$ metric in the extremal case and in the decoupling limit:
$r \ra 0, \alpha' \ra 0$, such that $U \equiv { r \over \alpha'}$ is kept 
fixed, gives:
 \eqn\demet{\eqalign
{& ds^{2}=\alpha'({U^{2} \over l^{2}}(-dt^{2}+
dx^{5~2})+{l^{2} \over U^{2}}dU^{2}+l^{2}d\Omega_{3}^{2}+
\sqrt{Q_{1} \over Q_{5}}
dx^{i}dx^{i}),\cr
}} 
where $x^{i}, i=5..9$ have $2 \pi$ period, and 
$l^{2}= g_{6} (Q_{1}Q_{5})^{1 \over 2}$. 
\par
Following $\malda$ we require the constants to be such that the radius of the
sphere to be one: $\alpha' l^{2}=1$. In this limit the volume of $T^{4}$  
shrinks to zero, justifying the use of $D=6$ supergravity coming from KK
 reduction of the $10D$ theory and the constant appearing in the
$D=6$ action is given by:
\eqn\ctrst{
{1 \over {g_{6}^{2} \alpha'^{2}}} \ra  N=Q_{1}~Q_{5}
}
From now on, we use such units that the overall constant in front
of the CFT action is only $N$, besides $l$ dependence. For each chiral primary
field in supergravity there is a corresponding one in CFT and in our case
the correspondence is : $s^{r}\ra {\it O}^{r}$ and $\sigma \ra {{\it O}_{0}}$.
The nonzero two point functions are then:
\eqn\tpfu{\eqalign
{ & \langle {\it O}^{r~I}(x){\it O}^{s~I_{1}}(y) \rangle=
N{l^{2}(l-1) \over 2^{l-2}}{\delta^{rs}\delta^{II_{1}} \over |x-y|^{2l}},\cr
  & \langle {\it O}_{0}^{I}(x){\it O}_{0}^{I_{1}}(y) \rangle=
N{l^{2}(l-1)^{2} \over 2^{l-2}(l+1)}{\delta^{II_{1}} \over |x-y|^{2l}}.\cr
}}
The two-point functions are normalized by multiplying the fields by 
appropriate factors: 
${\it O}^{I~r}$ by $({2^{l-2} \over N l^2(l-1)})^{1 \over 2}$ and
${\it O}_{0}^{I}$ by ${2^{l-2}(l+1) \over N l^2(l-1)^2}^{1 \over 2}$. Then 
the non-zero constants appearing in the three-point functions are given as:
\eqn\thrpfu{\eqalign
{& \langle {\it O}_{0}^{I} {\it O}^{r~I_{1}} {\it O}^{s~I_{2}} \rangle=
{\delta^{rs} \over N^{1 \over 2}}
{(l_{1}-1)^{1 \over 2}(l_{2}-1)^{1 \over 2} \over 
(l+1)^{1 \over 2}}\langle \Omega^{I} \Omega^{I_{1}}\Omega^{I_{2}}\rangle, \cr
& \langle {\it O}_{0}^{I} {\it O}_{0}^{I_{1}} {\it O}_{0}^{I_{2}} \rangle= 
{1 \over 6N^{1 \over 2}}{{l^2+l_{1}^2+l_{1}^2-2} \over 
(l+1)^{1 \over 2}(l_{1}+1)^{1 \over 2}(l_{2}+1)^{1 \over 2}}\langle \Omega^{I} \Omega^{I_{1}}\Omega^{I_{2}}\rangle, \cr
}}
where $\langle \Omega^{I} \Omega^{I_{1}}\Omega^{I_{2}}\rangle$ keeps track of
the $SO(4)$ Clebsch-Gordon coefficients, coming from spherical harmonics, 
in the structure constants.

\newsec{Conclusions}

In this paper, we gave the calculation of three point interactions of
chiral primaries in 6D supergravity. Deriving the interactions, we used 
nontrivial redefinitions of the fields which solved the linear equation
of motion for chiral primaries. The reasons behind these redefinitions are  
still not clear. The justification in our case is to use them to
remove unwanted derivative terms in the second order equations of motion. 
The appearance of the derivative terms shows in a way the lack in understanding
of supergravity compactified on spheres. It is after removing the derivative 
terms that we deal with a standard field theory in three dimensions. It would 
be very interesting to find a less technical principle to justify them 
in all orders. A step in this direction was made in $\nastase$ for a different 
case than ours and for a much smaller number of fields, namely those appearing 
in a consistent sphere truncation of the corresponding supergravity. 
 
\par
In section 4, we derived the three point interactions corresponding
to the expected conformal field theory in two dimensions.  If we compare 
results presented at the end of section 4 with the three point 
interactions found in the free orbifold CFT $\mijeram$ there are similarities,
but they are different overall. The similarities show that it is still possible
to expect a gravitational theory to be equivalent to the orbifold CFT, but the
question is which one. On the other side the differences are expected since 
the two descriptions do not sit at the same point in moduli space. A 
relevant role in this comparison is played by the dynamical  symmetry 
groups: on the gravity side there is a manifest $SO(5)$ for $T^4$ or 
$SO(21)$ for $K3$, symmetry at the lagrangian level. 
The backround we used, respects this symmetry and so does the expression for 
three point interactions derived in section 4. In the CFT only 
the $SO(4)$ or $SO(20)$ subgroup is present. At present, the dynamical mechanism
by which this is enlarged is not known. To achieve an exact comparison one 
should perform analogous computations in a different gravitational background 
or equivalently have a deformation of the CFT. This interesting question is 
left for future studies.

\bigskip

 \noindent{\bf Acknowledgements:}
 I am gratefull to Antal Jevicki for many helpfull disscussions and for his assistance
 in the preparation of this manuscript. I am also happy to acknowledge instructive discussions 
 with Sanjaye Ramgoolam and Radu Tatar. 
 \par
 I gratefully acknowledge a fellowship  from the Galkin Foundation.

 \bigskip
 \bigskip

\newsec{Appendix 1}
We give in this appendix detailed formulas of first 
( upper index $(1)$ ) and
second (upper index $(2)$) order variation of the $D=6$ fields used in text.
\par
We consider first the metric, the Christoffel symbols and
the curvature tensor:
\eqn\etrtr{\eqalign
{ & g^{MN}=\bar{g}^{MN}-h^{MN}+h^{2~MN}+.., \cr
 &\Gamma_{MN}^{P}=\bar{\Gamma}_{MN}^{P}+\Gamma_{MN}^{(1)~P}
 +\Gamma_{MN}^{(2)~P}+..,\cr
 &\Gamma_{MN}^{(1)~P}= {1 \over 2}
        (h^{~~P}_{M;N}+h^{~~P}_{N;M}-h^{~~~~~P}_{MN;}), \cr
&\Gamma_{MN}^{(2)~P}= {1 \over 2}h^{PR}
        (h_{RM;N}+h_{RN;M}-h_{MN;R}), \cr
  & R_{MN}=\bar{R}_{MN}+
(\Gamma_{MN;P}^{(1)P}-\Gamma_{MP;N}^{(1)P})+
(\Gamma_{MN;P}^{(2)P}-\Gamma_{MP;N}^{(2)P}+
 \Gamma_{RP}^{(1)~P}\Gamma_{MN}^{(1)~R}- \cr
  &~~~ -\Gamma_{RN}^{(1)~P}\Gamma_{MP}^{(1)~R})+.., \cr
}}
We are intersted only when $M=a, N=b$ and when the metric 
fluctuation are parametrized like in $\parm$ and the results is:
\eqn\mrtt{\eqalign
{& \bar{R}_{ab}=2\bar{g}_{ab}, \cr  
 & R^{(1)}_{ab}=\bar{g}_{ab}(-{1 \over 2}(\lpa N+ \lps N)
-{1 \over 6} \lps (3M+N))- {1 \over 2}(3M+N)_{;(ab)},\cr
  & R^{(2)}_{ab}=\bar{g}_{ab}{1 \over 12}(4 (N^{2})_{;c}^{~~c}+6 (N_{,\mu} h^{\mu \nu})_{;\nu}+(h^{2~\mu}_{\mu})_{;c}^{~~c}-
h^{\mu \nu}_{~~;c}h_{\mu \nu;}^{~~c}-N_{,c}N_{,}^{~c})+\cr
&~~~+{1 \over 4}((N^{2}+h^{2~\mu}_{\mu})_{;(ab)}-h^{\mu \nu}_{~~;(a}h_{\mu \nu;b)}-N_{,(a}N_{,b)}). \cr
}}
We should also compute $\etrtr$ for the case $M=\mu, N=a$, but because the only thing
we extract from this is $h'$, needed only in linearized approximation we refer to $\sezgin$.
\par
Next we consider the forms and the scalars given in $\fluctsec$ and we need 
only  
the fluctuations in Hodge duality for forms and in the covariant derivative for scalars, 
due to metric fluctuations:
\eqn\hdua{\eqalign
{ & *H_{MNP}={1 \over 3!} \epsilon_{MNP}^{~~~~QRS}
((1+{1 \over 2} h_{M}^{~M}+{1 \over 4}({1 \over2}
(h_{M}^{~M})^{2}+h_{~M}^{2~~M}))H_{QRS}+ \cr
  &~~~+3(-h_{Q}^{~Q'}+h_{~Q}^{2~~Q'}H_{Q'RS}+6h_{Q}^{~Q'}h_{R}^{~R'}H_{Q'R'S})+.., \cr
  & D_{M}P_{N}^{ir}={1 \over \sqrt{2}}( \bar{\nabla_{M}}\partial_{N}\phi^{ir}+
             \Gamma_{MN}^{(1)~P}\partial_{P}\phi^{ir})+..,\cr
}}
Using the definitions for fields, we obtain in a straightforward manner the 
equations given in text.
\newsec{Appendix 2}
We list in this appendix the formulas we need to know about spherical 
harmonics on $S^{3}$. We define the scalar harmonics similar to
$\leeseiberg$, namely:
\eqn\dhar{
Y^{I}\equiv {1 \over r^{l}} x^{(i_{1}}..x^{i_{l})}\Omega^{I}_{i_{1}..i_{l}},
}
where $x^{i}, i=1..4$ are four flat coordinates, 
$r=\sqrt{x_1^{~2}+..+x_4^{~2}}$ and the sphere $S^3$ is embedded as $r=1$. The
pharantheses $(..)$ means like anywhere in text, the symmetrized and 
traceless tensor and $\Omega^{I}$'s form 
an orthonormal basis of constant symmetrized and traceless tensors. 
From $\dhar$ is clear that $Y^{I}$ does not depend on $r$, but only on sphere indices. We also need:
\eqn\prop{\eqalign
{& \lps Y^{I}=-l(l+2)Y^{I}, \cr
 & \lps (\partial_{a}Y^{I})=(-l(l+2)+2)\partial_{a}Y^{I}. \cr
}}
We define $\Delta=l(l+2)$. Considering the basic constants we need to compute
and determine all the other spherical harmonic related constants in term of 
these:
\eqn\ctos{\eqalign
{& C(I) \equiv \int_{S^3} Y^{I} Y^{I}, \cr
 & \ca \equiv \int_{S^3} Y^{I}Y^{I_{1}}Y^{I_{2}}. \cr
 }}
 We can now prove using eventually simple insertions of $\lps$ in the 
equations $\ctos$ the following identities:
\eqn\ctost{\eqalign
{& \int_{S^3} Y^{I} Y^{I_{1}}=C(I)\delta^{I,I_{1}}, \cr
 & \int_{S^3} \partial_{a}Y^{I} \partial^{a} Y^{I_{1}}=C(I)\delta^{I,I_{1}}\Delta, \cr
 & \int_{S^3} \nabla_{(a}\nabla_{b)}Y^{I} \nabla^{(a}\nabla^{b)}Y^{I_{1}}=C(I)\delta^{I,I_{1}} {2 \over 3}\Delta(\Delta-3). \cr
 }}
and
\eqn\ctostr{\eqalign
{ & \int_{S^3} Y^{I}\nabla_{a}Y^{I_{1}}\nabla^{a}Y^{I_{2}}={1 \over 2}(\Delta_{1}+\Delta_{2}-\Delta)\ca, \cr
  & \int_{S^3} \nabla^{(a}\nabla^{b)}Y^{I}\nabla_{(a}Y^{I_{1}}\nabla_{b)}Y^{I_{2}}=
  {1 \over 4}(\Delta_{1}+\Delta_{2}-\Delta-4)(\Delta_{1}+\Delta_{2}-\Delta)\ca, \cr
  & \int_{S^3} Y^{I}\nabla_{(a}\nabla_{b)}Y^{I} \nabla^{(a}\nabla^{b)}Y^{I_{1}}=
  ({1 \over 4}(\Delta_{1}+\Delta_{2}-\Delta-4)(\Delta_{1}+\Delta_{2}-\Delta)-{1 \over 3}
  \Delta_{1}\Delta_{2})\times \cr
  & ~~~ \times  \ca. \cr
  }}
 We compute the constants defined in $\ctos$ using gaussian integrals in $x$-space and we 
 obtain the following:
 \eqn\ctosdet{\eqalign
 {& C(I)={{2 \pi^{2}} \over {2^{l} (l+1)}}, \cr
  & \ca ={{2 \pi^{2}} \over 2^{{\Sigma \over 2}}}{{l! l_{1}! l_{2}!} \over 
  {{\alpha \over 2}! {\alpha_{1} \over 2}! {\alpha_{2} \over 2}! ({\Sigma \over 2}+1)!}}
  \langle \Omega \Omega^{I_{1}}\Omega^{I_{2}}\rangle, \cr
  }}
  where the last factor represents the possible contractions of all the indices of the 
  three normalized tensors.
  \par
 After working everything out we write a $D=3$ action responsible for the reduced (on $S^3$) 
 equations of motion. It is the spirit of the correspondence to relate the higher dimensional
 vertices to correlation functions in a conformal field theory one lower dimension. The explicit
correspondence was given in $\freedmanmathur$ and we repeat here the results applicable to our case. 
The two and three point correlation functions in a conformal field theory are completely
determined up to a constant. We give below only the constants assuming that the linearized
equations are those of correponding chirals ( $\eta$ is a constant coming from higher 
dimensional supergravity and $\delta=l(l-2)$):
\eqn\corras{\eqalign
{ & {\eta \over 2}((\partial s)^2+\delta~ s^2) \ra \eta{l(l-1) \over \pi}, \cr
  & \eta~ s s_{1} s_{2} \ra \eta {{({\alpha \over 2}-1)! {(\alpha_{1} \over 2}-1)!
   ({\alpha_{2} \over 2}-1)! ({\Sigma \over 2}-1)!} \over {2 \pi^2 (l-2)! (l_{1}-2)!
   (l_{2}-2)!}}. \cr
  }}

\listrefs

\end